\documentclass[twoside]{article}
\usepackage[english]{babel}
\usepackage{fancyhdr}
\usepackage[a4paper,top=2cm,bottom=2cm,left=3cm,right=3cm,marginparwidth=1.75cm]{geometry}

\usepackage[table, xcdraw, dvipsnames]{xcolor}
\usepackage{cite}
\usepackage{amsmath,amssymb,amsfonts}
\usepackage{algorithmic}
\usepackage{graphicx}
\usepackage{textcomp}
\usepackage{pgfplots} 
\usepackage{minted}
\usepackage{listings}
\usepackage{tabularx}
\usepackage{multirow}
\usepackage{booktabs}
\usepackage{array}
\usepackage{pdflscape}
\usepackage{adjustbox}
\usepackage{svg}
\usepackage{circledsteps}
\usepackage{wasysym}
\definecolor{light-gray}{gray}{0.95}
\usepackage{pgf-pie}  
\usepackage[T1]{fontenc}
\usepackage{tgadventor}
\usepackage{tcolorbox}
\tcbuselibrary{breakable}
\usepackage{caption}
\usepackage[outline]{contour}
\usepackage{array}
\usepackage{authblk}
\usepackage{pgf-pie}
\usepackage{adjustbox}
\usepackage{tikz}
\usetikzlibrary{arrows.meta, positioning, bending, calc}
\usepackage{subcaption} 
\usepackage{placeins}

\usepackage{hyperref}

\newcommand{\pie}[1]{%
\begin{tikzpicture}
 \draw (0,0) circle (1ex);\fill (1ex,0) arc (0:#1:1ex) -- (0,0) -- cycle;
\end{tikzpicture}%
}

\newcolumntype{P}[1]{>{\centering\arraybackslash}p{#1}}

\definecolor{mycolor}{RGB}{9,99,125}

\makeatletter
\newcommand\footnoteref[1]{\protected@xdef\@thefnmark{\ref{#1}}\@footnotemark}
\makeatother

\newcommand{\tool}{\textsc{CFIghter}}

\lstdefinestyle{code}{
    frame=lines,
    numbers=left,
    xleftmargin=20pt,
    basicstyle=\ttfamily\footnotesize,
    numberstyle=\tiny\color{gray},
    keywordstyle=\color{blue},
    commentstyle=\color{gray},
    stringstyle=\color{orange},
    breaklines=true,
    showstringspaces=false,
    tabsize=4
}

\lstdefinelanguage{Assembler}{
  morekeywords={mov,add,sub,cmp,jne,je,jmp,call,ret,push,pop,lea,ud2},
  morecomment=[l]{;},
  morestring=[b]",
  sensitive=true
}

\lstdefinelanguage{IR}{
  morekeywords={constant, i32, load, call, switch},
  morecomment=[l]{;},
  morestring=[b]",
  sensitive=true
}

    \definecolor{neoncarrot}{rgb}{1.0, 0.64, 0.26}
    \definecolor{rosevale}{rgb}{0.67, 0.31, 0.32}
    \definecolor{olivine}{rgb}{0.6, 0.73, 0.45}
    \definecolor{ao}{rgb}{0.0, 0.5, 0.0}

\definecolor{light-gray}{gray}{0.95}

\newcolumntype{R}[1]{>{\raggedleft\let\newline\\\arraybackslash\hspace{0pt}}m{#1}}
\newcolumntype{L}[1]{>{\raggedright\let\newline\\\arraybackslash\hspace{0pt}}m{#1}}

\title{\tool{}: Automated Control-Flow Integrity Enablement and Evaluation for Legacy C/C++ Systems}
\author{Sabine Houy}
\author{Bruno Kreyssig}
\author{Alexandre Bartel}
\affil{Department of Computing Science, Ume\aa~University}
\date{}

\begin{document}
\maketitle

\begin{abstract}
Compiler-based Control-Flow Integrity (CFI) offers strong forward-edge protection but remains challenging to deploy in large C/C++ software due to visibility mismatches, type inconsistencies, and unintended behavioral failures. 
We present \tool{}, the first fully automated system that enables strict, type-based CFI in real-world projects by detecting, classifying, and repairing unintended policy violations exposed by the test suite. 
\tool{} integrates whole-program analysis with guided runtime monitoring and iteratively applies the minimal necessary adjustments to CFI enforcement only where required, stopping once all tests pass or remaining failures are deemed unresolvable.

We evaluate \tool{} on four GNU projects. 
It resolves all visibility-related build errors and automatically repairs 95.8\% of unintended CFI violations in the large, multi-library \texttt{util-linux} codebase, while retaining strict enforcement at over 89\% of indirect control-flow sites. 
Across all subjects, \tool{} preserves strict type-based CFI for the majority of the codebase without requiring manual source-code changes, relying only on automatically generated visibility adjustments and localized enforcement scopes where necessary.
These results show that automated compatibility repair makes strict compiler CFI practically deployable in mature, modular C software.
\end{abstract}

\section{Introduction}
Memory corruption has a long history of causing critical vulnerabilities, including the infamous EternalBlue~\cite{eternalblue}, DirtyPipe~\cite{dirtypipe}, and a plethora of recent browser-engine exploits\footnote{including \href{https://nvd.nist.gov/vuln/detail/CVE-2025-10585}{CVE-2025-10585}, \href{https://nvd.nist.gov/vuln/detail/CVE-2024-5493}{CVE-2024-5493}, and \href{https://nvd.nist.gov/vuln/detail/cve-2024-9680}{CVE-2024-9680}}. 
Even with the emergence of memory-safe languages such as Rust, C/C++ code still forms the bedrock of modern operating systems, and widely used software \cite{houy2025twenty,becker2024sok}. 
This makes mitigation techniques for memory-unsafe languages particularly relevant.

Compiler-based \textit{Control-Flow Integrity} (CFI) has emerged as an effective mechanism to mitigate control-flow hijacking attacks caused by memory corruption vulnerabilities in C and C++ programs~\cite{abadi2009control,tice2014enforcing}. 
CFI enforces restrictions on indirect control transfers, such as calls via function pointers or virtual dispatch. 
Each transfer is allowed to target only a statically determined and restricted set of legitimate functions. 
Although practical implementations inevitably rely on over-approximation of valid targets, CFI significantly raises the bar for code-reuse attacks.
Modern compiler infrastructures such as LLVM/Clang provide modular CFI implementations that apply fine-grained~\cite{LLVMdoc,tice2014enforcing} runtime checks with low overhead~\cite{LLVMdoc,LLVMDesigndoc}, making CFI suitable for integration into production-grade software.

However, despite significant progress in compiler-based implementations~\cite{androidcfi,becker2024sok,houy2025twenty}, deploying CFI in large, real-world software projects remains challenging. 
The main difficulty is not the instrumentation process itself, but the \textit{semantic mismatch} between CFI's security policies and the diverse, often unconventional behaviors allowed by the C/C++ language and its ecosystem~\cite{firefoxcfi,chromiumcfi,xu2019confirm,houy2024lessons,houy2025twenty}. 
Many C and C++ projects use constructs that conflict with the assumptions of strict CFI enforcement, such as flexible function pointer usage, custom polymorphism models, dynamic loading, or non-standard casting conventions~\cite{houy2025sok}. 
From the perspective of the CFI runtime, these patterns appear as invalid or unsafe control transfers, even though they are correct and intentional according to the program's semantics or the C standard. 
As a result, CFI instrumentation often triggers unintended policy violations, leading to \textit{Illegal instruction} (\texttt{SIGILL}) exceptions and runtime crashes during testing or execution. 
These false positives make CFI adoption brittle and impractical without extensive manual debugging, source modification, or policy relaxation.

To address this fundamental compatibility challenge, we design the novel tool \tool{}. It automatically detects, classifies, and mitigates unintended CFI policy violations. 
\tool{} preserves LLVM's enforcement semantics while applying targeted fixes that confine necessary exceptions to specific, verified cases and maximize protected code coverage. 
It analyzes build and test outputs to determine why enforcement failures occur. 
\tool{} distinguishes between two leading causes of enforcement failures. 
The first category includes technical build and linkage errors, such as missing symbols or unresolved references introduced by CFI instrumentation. 
The second category consists of semantic incompatibilities, where legitimate program behavior, such as indirect calls across modules or dynamic interactions, conflicts with CFI's static assumptions. 
When such cases are detected, \tool{} automatically updates its ignorelist for semantic policy exceptions or modifies source-level visibility attributes where necessary to restore linkage consistency. 
Through this iterative self-healing process, \tool{} progressively increases CFI coverage while maintaining the program's functionality, although minor deviations in test outcomes may still occur for certain functional edge cases.

To ensure compatibility across different enforcement modes, \tool{} supports all seven modular LLVM forward-edge CFI variants~\cite{LLVMdoc}. 
This comprehensive support enables \tool{} to consistently detect and repair policy violations across diverse instrumentation types, ensuring its adaptation process remains effective regardless of the specific CFI variant or compiler configuration.

In summary, this paper introduces \tool{}, an automated framework that addresses the compatibility and integration challenges that currently limit the adoption of compiler-based Control-Flow Integrity in practice. 
Building on automated analysis and iterative repair, this work makes the following contributions:
\begin{itemize}
    \item We identify three main classes of challenges: (i) semantic incompatibilities, (ii) structural and toolchain constraints, and (iii) integration overhead. 
    \item We develop \tool{}, a novel framework that combines static binary inspection with dynamic process tracing to detect and repair CFI-related build and test-time failures. \tool{} is publicly available, see Section~\ref{sec:data}.
    \item We demonstrate \tool{}'s effectiveness on large, multi-library legacy software, automatically enabling strict CFI enforcement in projects that previously failed under compiler-based CFI. 
    We evaluate \tool{} on four open-source legacy projects and find that it automatically resolves all visibility-related build errors and $\sim$95.8\% of unintended policy violations, resulting in a CFI coverage of $\sim$89\% indirect call sites. 
\end{itemize}
 
Across multiple LLVM forward-edge variants, \tool{} substantially increases CFI coverage and reduces the manual effort required for integration. 
These results demonstrate that automated compatibility repair can make compiler-enforced CFI practically deployable in structurally diverse, real-world C/C++ software.

\section{Background}
In this Section we first introduce the general principle of how compiler-based CFI protects code from undesired control flow redirection, followed by the seven variants LLVM provides for CFI enforcement. 
Then, we summarize the main hurdles for applying CFI to legacy codebases, including link-time optimization visibility (Section \ref{sec:lto}) and three main integration challenges (Section \ref{sec:challenges}).

\subsection{Control-Flow Integrity}
Compiler-based \textit{Control-Flow Integrity} (CFI) is a mitigation technique that restricts the targets of indirect control transfers to a set of legitimate destinations determined at compile time~\cite{abadi2009control,tice2014enforcing}.  
It was designed to prevent \textit{control-flow hijacking} attacks, in which an adversary exploits memory corruption vulnerabilities to redirect execution to malicious code or unintended program regions.  
Typical examples are \textit{return-oriented programming} (ROP)~\cite{zhang2015control,davi2014hardware} and \textit{jump-oriented programming} (JOP)~\cite{christoulakis2016hcfi}.  
In these attacks, an adversary chains together existing instructions, so-called \textit{gadget chains}, to perform arbitrary computation without injecting code. 
CFI mitigates such abuse by restricting indirect jumps, function-pointer calls, and virtual dispatches to compiler-approved targets.  
This constraint significantly raises the effort required for reliable exploitation~\cite{houy2025sok}. 

In LLVM/Clang, CFI~\cite{LLVMdoc,LLVMlto} is implemented by associating each function or virtual call target with type metadata and inserting runtime checks before every indirect call or virtual dispatch.  
At runtime, these checks verify that the actual call target matches the expected dynamic type, preventing the program from transferring control to unintended code. 
Constructing this mapping of valid call targets requires a whole-program view of the codebase, including all functions and their relations.  
To obtain such visibility, LLVM's CFI requires \textit{Link-Time Optimization} (LTO) to merge all intermediate representations (IR) during the final link stage, as discussed in Section~\ref{sec:lto}. 

CFI mechanisms can be broadly divided into \textit{forward-edge} and \textit{backward-edge} protection. 
For-ward-edge CFI constrains indirect calls, such as function-pointer invocations, virtual dispatches, and computed jumps.  
Backward-edge CFI, by contrast, protects \texttt{return} instructions by ensuring that return addresses on the stack have not been altered, often through techniques like shadow stacks~\cite{dang2015performance,burow2019sok, LLVMscs}.  
This work focuses exclusively on forward-edge CFI, as these protections depend heavily on compiler-level metadata, symbol visibility, and linkage information.

\subsubsection{Example of Compiler-Based Instrumentation}
Listing~\ref{lst:cfi_example_cpp} illustrates the principle of compiler-based CFI  for one of its variants (indirect calls). 
\begin{figure}[!ht]
\centering
\begin{lstlisting}[style=code,language=C++]
typedef void (*func_t)(void);

void target() { ... }

void caller(func_t foo) {         
    // indirect call, instrumented by CFI
    foo();
}

int main() {
    caller(target);
}
\end{lstlisting}
\caption{Simplified example of forward-edge CFI instrumentation in LLVM.}
\label{lst:cfi_example_cpp}
\end{figure}
The compiler associates a type identifier with each function signature (line~1), when compiled with Clang using \textit{LTO \& hidden-visibility} and indirect call CFI protection \texttt{cfi-icall}. 
It then inserts a runtime type check before the indirect call (line~7).  
This results in the assembly instructions in Listing~\ref{lst:cfi_example_asm}. 
\begin{figure}[!ht]
\begin{lstlisting}[style=code, language=Assembler]
__caller:
    ; load function pointer argument
    mov   rax, [rdi]             
    ; load expected type identifier
    mov   rcx, typeid@target   
    ; compare expected vs. actual type
    cmp   [rax + typeid@off], rcx 
    ; abort if type mismatch detected
    jne   __cfi_check_fail       
    ; perform verified call
    call  rax          

__cfi_check_fail:
    ; causing SIGILL -> crash
    ud2
\end{lstlisting}
\caption{Simplified pseudo-assembly showing CFI type checking before an indirect call.}
\label{lst:cfi_example_asm}
\end{figure}

At runtime, the CFI instrumentation verifies that the function pointer passed to \texttt{caller()} (line~5 in Listing~\ref{lst:cfi_example_cpp}) refers to a function whose dynamic type matches the expected signature (lines~5 \& 7 in Listing~\ref{lst:cfi_example_asm}). 
If the comparison fails (line~7 in Listing~\ref{lst:cfi_example_asm}), control is transferred to the failure handler, which terminates execution with an \textit{Illegal instruction} (\texttt{ud2} instruction) exception (line 15).
Otherwise, the verified target function is invoked normally (line~11). 
This runtime validation ensures that even if an attacker corrupts a function pointer, the program cannot redirect execution to an arbitrary address without triggering a CFI violation.

\subsubsection{Granularity of Enforcement}
CFI mechanisms differ in the granularity of their enforcement policies.  
\textit{Coarse-grained} schemes group multiple indirect call targets into broad equivalence classes (for example, all functions within a module or sharing a prototype), thereby permitting more valid targets but offering weaker protection~\cite{carlini2015control}.  
\textit{Fine-grained} or \textit{strict} CFI, in contrast, constructs precise equivalence classes based on type metadata or control-flow graphs, allowing only the exact set of targets that match the intended dynamic type~\cite{tice2014enforcing}.  
LLVM's forward-edge CFI variants enforce type-consistent control transfers at each call-site, providing fine-grained protection~\cite{LLVMdoc}.  
Yet, this strict enforcement can interact adversely with common C/C++ patterns and build configurations, leading to practical compatibility challenges~\cite{firefoxcfi,chromiumcfi,xu2019confirm,houy2024lessons,houy2025twenty}. 
However, even fine-grained enforcement is subject to fundamental precision limits, as discussed in Section~\ref{sec:cfi_limitations}.

\subsubsection{LLVM Forward-Edge CFI Variants}
LLVM implements several modular forward-edge variants, each targeting a specific category of indirect control transfers or unsafe casts, as summarized in Table~\ref{tab:variants}.  
\begin{table}[ht!]
    \centering
    \caption{LLVM Forward-Edge CFI Variants~\cite{LLVMdoc}}
    \label{tab:variants}
    \begin{tabular}{rp{0.5\linewidth}}
    \toprule
        \textsc{CFI Variant}        & 
        \textsc{Description}        \\
    \midrule
        \rowcolor{light-gray}
        \texttt{cfi-icall}          & 
        Indirect call of a function whose dynamic type does not match the expected function type. \\
        \texttt{cfi-vcall}          & 
        Virtual call via an object whose virtual pointer (\texttt{vptr}) refers to an object of the wrong dynamic type. \\
        \rowcolor{light-gray}
        \texttt{cfi-nvcall}         & 
        Non-virtual call via an object whose \texttt{vptr} refers to an object of the wrong dynamic type. \\
        \texttt{cfi-mfcall}         & 
        Indirect call through a member function pointer that targets a function of the wrong dynamic type. \\
        \rowcolor{light-gray}
        \texttt{cfi-cast-strict}    & 
        Enables strict checking for all invalid base-to-derived casts, disabling the relaxed behavior normally permitted for layout-compatible classes. \\
        \texttt{cfi-derived-cast}   & 
        Detects invalid base-to-derived casts where the destination object has an unexpected dynamic type. \\
        \rowcolor{light-gray}
        \texttt{cfi-unrelated-cast} & 
        Detects invalid casts from \texttt{void*} or other unrelated types to an object of the wrong dynamic type. \\ 
    \bottomrule
    \end{tabular}
\end{table}

These variants collectively enforce type- and object-consistent control transfers in C and C++ programs~\cite{LLVMDesigndoc}.  
They target indirect function calls, virtual dispatches, and pointer casts that could otherwise be abused to redirect control flow.  
An LLVM CFI integration framework should support all seven CFI variants. 
This ensures that, across diverse CFI enforcement configurations and compiler settings, the automated repair process effectively solves the main integration challenges.

\subsection{Link-Time-Optimization \& Visibility}\label{sec:lto}
LLVM CFI's requirement for Link-Time Optimization (LTO) is one of the main impediments to adopting CFI in legacy codebases.
LTO enables the compiler to perform cross-module analysis, build a complete control-flow graph, and instrument the program consistently across all translation units.  
LLVM's forward-edge CFI relies on whole-program analysis at link-time. Thus, CFI builds must enable LTO with the flag \texttt{-flto}.  
Equally important, the compiler will only emit CFI checks for functions and classes when it can infer \textit{hidden LTO visibility}.  
In practice, this means building with \texttt{-fvisibility=hidden} so that targets are not overridden across shared libraries and are fully visible within the LTO unit~\cite{LLVMdoc,LLVMlto}.  
If \texttt{default} (\textit{public}) visibility is used, many entities are treated as potentially replaceable at link or runtime, and CFI checks for those entities are disabled~\cite{LLVMdoc}.  

LLVM's CFI requires hidden visibility (similar to \texttt{private}) because the soundness of its control-flow checks depends on complete, \textit{closed-world} knowledge of all valid indirect call targets.  
In dynamically linked environments, symbols with default visibility may be overridden or extended by other shared objects through mechanisms such as symbol interposition or subclassing across shared libraries.  
This violates the compiler's assumption of a fixed control-flow graph, making static enforcement of CFI unsound.  
By restricting symbols to hidden visibility, the compiler and linker can guarantee that no external code introduces additional call targets, allowing LLVM's link-time analysis to construct a complete and immutable control-flow graph. 

LLVM additionally provides \textit{cross-DSO CFI} to enforce control-flow protection across shared libraries.  
In this mode, each dynamic shared object (DSO) must be compiled with \textit{cross-DSO} support and registers its valid call targets with the runtime through the \texttt{\_\_cfi\_check} interface at load time~\cite{LLVMdoc,LLVMDesigndoc,crossDSO}.  
Although this mechanism maintains functionality in dynamically linked environments, it assumes complete control over the build of all participating libraries. 
This introduces runtime registration overhead and nondeterminism in the set of valid targets.  
Because these constraints conflict with \tool{}'s goal of deterministic, whole-program analysis, the framework operates exclusively under the standard single-LTO model.

These requirements interact with real-world projects in two ways.  
First, enforcing hidden visibility can surface missing exports and unresolved references during LTO.  
Second, some legitimate cross-module patterns, such as callbacks that cross shared library boundaries, require that selected symbols remain externally accessible.  
In such cases, developers typically maintain a \textit{hidden-by-default} policy but selectively mark necessary APIs or symbols with \texttt{visibility("default")} to preserve intended linkage while retaining CFI enforcement elsewhere~\cite{redhatVisibility}.  
This approach maintains correct linkage behavior while ensuring that LLVM can still emit CFI checks for all remaining hidden symbols. 

Our goal is to automate this approach by iteratively building with LTO and hidden visibility, analyzing errors that arise from these settings, and then selectively relaxing visibility on the minimal set of functions required for correct cross-module behavior.   
The result is high CFI coverage under the required \textit{LTO \& hidden-visibility} configuration, with targeted exceptions rather than global relaxation.

\begin{table*}[hb!]
\centering
\caption{Mapping of practical CFI challenges to their causes, effects, and \tool{} components addressing them.}
\label{tab:cfi_challenges}
    \begin{tabular}{p{0.11\linewidth}p{0.26\linewidth}p{0.26\linewidth}p{0.26\linewidth}@{}}
    \toprule
        \textbf{Challenge} & \textbf{Root Cause} & \textbf{Example Impact} & \textbf{Addressed by \tool{}} \\ 
    \midrule
        \rowcolor{light-gray}
        Semantic incompatibilities (Section~\ref{sec:semantic_challengescfi}) &
        Type- or CFG-based CFI policies conflict with flexible language constructs such as dynamic linking, custom casts, or plugin interfaces. &
        Legitimate indirect calls are misclassified as policy violations (\texttt{SIGILL}); compilation or tests crash. &
        \ref{sec:analysis_phase}~\textit{Analysis Phase} -~\Circled[inner color=white, fill color=black]{\textnormal{\textsf{B}}} -- classifies violations, identifies the root cause, and updates the ignorelist. \\
        
        Structural / toolchain constraints (Section~\ref{sec:structural_constraints}) &
        CFI requires LTO and hidden LTO visibility, which conflict with modular builds or exported library symbols. &
        Linking failures or missing instrumentation in shared-library builds. &
        \ref{sec:building_phase}~\textit{Building Phase} -~\Circled[inner color=white, fill color=black]{\textnormal{\textsf{A}}} -- automatically detects visibility errors and relaxes affected symbols selectively. \\
        
        \rowcolor{light-gray}
        Integration overhead (Section~\ref{sec:developer_overhead}) &
        Manual debugging, configuration tuning, and visibility management across multiple CFI variants. &
        High engineering cost and inconsistent deployment across projects. &
        \textsc{all} / ~\ref{sec:testing_phase}~\textit{Testing Phase} -~\Circled[inner color=white, fill color=black]{\textnormal{\textsf{C}}} -- automates validation and regression comparison between baseline and CFI builds. \\ 
    \bottomrule
    \end{tabular}
\end{table*}

\subsection{Challenges of Applying CFI to Existing Codebases}\label{sec:challenges}
While compiler-based CFI mechanisms provide substantial mitigation against control-hijacking attacks, applying them to large, legacy C and C++ codebases remains far from straightforward.  
Three main classes of challenges hinder adoption in practice. 
These challenges were derived from prior CFI deployment studies and evaluations~\cite{houy2024lessons,becker2024sok,houy2025twenty}, as well as our own preliminary attempts to apply LLVM's CFI to large codebases. 
We focus on three representative classes, not an exhaustive taxonomy.

\subsubsection{Semantic incompatibilities}\label{sec:semantic_challengescfi}
Many real-world projects rely on language features or design patterns that conflict with CFI's strict assumptions.  
Flexible use of function pointers, complex casting idioms, runtime plugin systems, and cross-language interfaces often result in indirect control transfers that are legitimate under the C/C++ standard but appear unsafe to the compiler or more precisely to the compiler's CFI component~\cite{carlini2015control,houy2025sok,evans2015control}.  
For example, non-standard type casts or dynamic linking may lead to valid function calls that CFI instrumentation misclassifies as policy violations.
\subsubsection{Structural and toolchain constraints}\label{sec:structural_constraints}
Numerous codebases employ modular build systems, only partially hidden symbols, or incremental linking.  
These practices conflict with CFI's requirement for whole-program visibility through link-time optimization and hidden LTO visibility.  
LLVM's CFI requires hidden LTO visibility to ensure that functions cannot be replaced at runtime by definitions in other shared libraries, allowing the compiler to emit precise type checks.  
However, this requirement conflicts with common software designs that export numerous symbols or rely on dynamic linking between shared objects~\cite{houy2025sok,becker2024sok,muntean2019analyzing}.  
When visibility is globally set to hidden, external interfaces may fail to link or become inaccessible to other components, even though CFI itself depends on this configuration for functionality.  
Projects with shared libraries or plug-in architectures, therefore, often face build failures or restricted functionality under strict CFI settings.  
\subsubsection{Integration overhead}\label{sec:developer_overhead}
The two challenges mentioned above can typically be solved only with significant manual effort. 
Developers must inspect build logs, trace linkage failures, and modify source code to adjust visibility attributes and maintain ignorelists.  
Such iterative debugging and configuration tuning significantly increase the cost of adoption, discouraging deployment in production environments~\cite{becker2024sok,houy2025twenty,tice2014enforcing,houy2024lessons}.

\section{Methodology}

\noindent Table~\ref{tab:cfi_challenges} summarizes the primary obstacles identified in existing CFI deployments, their underlying causes, and how each is addressed within \tool{}'s architecture.  
This mapping provides an overview of the relationship between the problem space and the design principles detailed in the following subsections.

\subsection{\tool{}'s Design}\label{sec:design}
\begin{figure*}[!t]
  \centering
  \includegraphics[width=\linewidth]{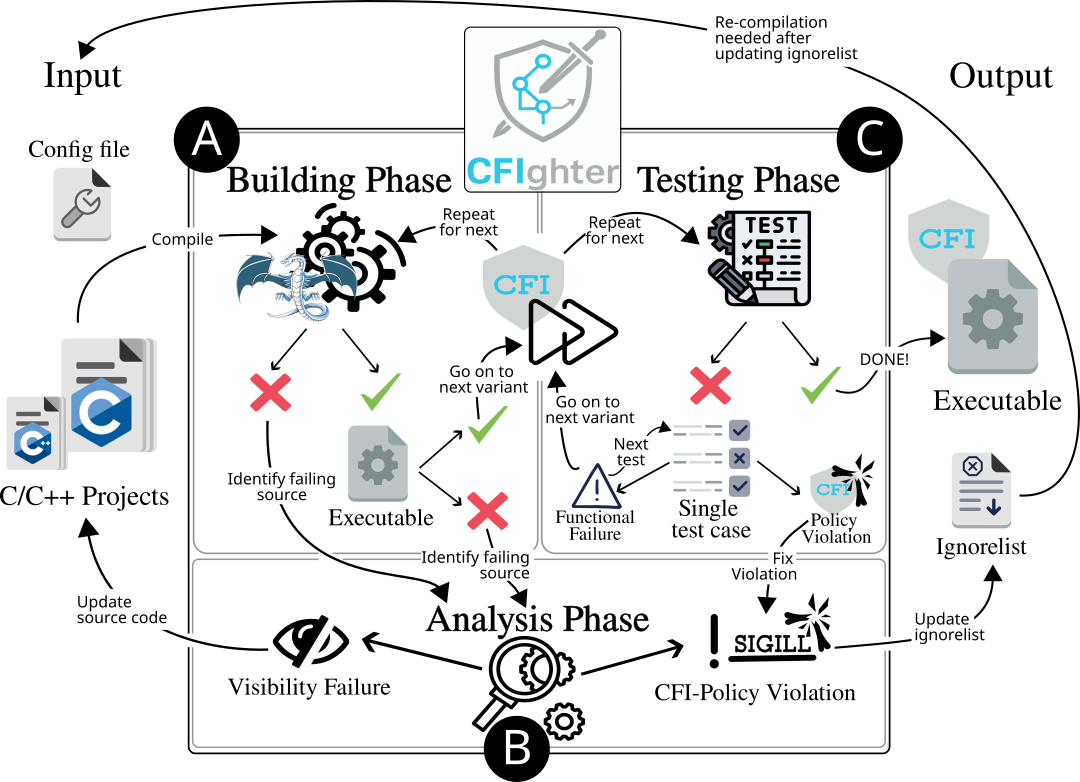}
\caption{\protect\tool's automated self-healing pipeline for adapting large C/C++ projects to LLVM's Control-Flow Integrity (CFI). 
    The framework operates in three phases: 
    the \textit{Building Phase} (\protect\Circled[inner color=white, fill color=black]{\textsf{A}}), 
    the \textit{Analysis Phase} (\protect\Circled[inner color=white, fill color=black]{\textsf{B}}), and 
    the \textit{Testing Phase} (\protect\Circled[inner color=white, fill color=black]{\textsf{C}}), which compares baseline and protected executions to iteratively repair and validate CFI enforcement.}
  \label{fig:pipeline}
\end{figure*}

\tool{} is an automated framework designed to detect and repair LLVM's CFI-related build and runtime failures in large-scale C/C++ projects. 
Its overarching goal is to apply CFI protection to as much code as possible without breaking the project's functionality. 
By automating the identification and resolution of both visibility and policy violations, \tool{} ensures that even large, structurally diverse legacy software can be instrumented with strong CFI guarantees while maintaining functionality. 
Figure~\ref{fig:pipeline} illustrates the overall workflow of \tool{}, which consists of three interconnected phases: the \textit{Building Phase}, \textit{Analysis Phase}, and \textit{Testing Phase}. 
Together, these phases form a self-healing pipeline that iteratively compiles, tests, diagnoses, and repairs the software until a stable and secure configuration is achieved. 

\subsubsection{Building Phase -~\Circled[inner color=white, fill color=black]{\textnormal{\textsf{A}}}}\label{sec:building_phase}
It is responsible for generating all executable targets of the target project. 
Using a configuration file that specifies the project root, build commands, and executables to be analyzed, \tool{} automatically compiles both the baseline and the CFI-instrumented versions of the codebase. 
Baseline builds are compiled without instrumentation to provide a reference for correct program behavior, while CFI builds are compiled with Clang/LLVM under different CFI policies.

During compilation, \tool{} captures the build output to detect undefined references, missing symbols, or linkage errors. 
Such failures typically correspond to visibility issues that arise when compiler instrumentation restricts function accessibility across modules. 
Whenever such issues occur, \tool{} transitions to the \textit{Analysis Phase} to diagnose the failure and automatically adjust the project's visibility attributes. 
This process repeats until all variants compile successfully, ensuring that CFI can be applied to the maximum number of components without breaking the build.

\subsubsection{Analysis Phase -~\Circled[inner color=white, fill color=black]{\textnormal{\textsf{B}}}}\label{sec:analysis_phase}

The analysis phase forms the adaptive core of \tool{} and is responsible for diagnosing and repairing all failures encountered during compilation or execution. 
In this phase \tool{} separates between two fundamentally different failure modes: (i) \textit{visibility failures} that prevent successful compilation and linking, and (ii) \textit{CFI-policy violations} that arise only at runtime when LLVM's forward-edge CFI rejects a legitimate indirect control-flow transfer. 
Treating these classes independently is essential: visibility failures must be resolved \textit{first}, since they block test execution and thereby preclude any meaningful runtime interpretation.

After compilation succeeds, \tool{} executes each binary or tests causing a CFI policy violation under \texttt{ptrace}\footnote{\noindent\url{https://man7.org/linux/man-pages/man2/ptrace.2.html}}, which provides fine-grained control-flow monitoring and signal interception (more technical details in Appendix~\ref{app:ptrace-unwinding}. 
At this stage, our goal is to detect the method invocations that cause CFI-induced crashes so that we can add them to the ignorelist and recompile without CFI.
On \texttt{x86\_64}, LLVM's CFI signals all forward-edge policy violations using a deterministic \texttt{SIGILL}\footnote{In contrast, on AArch64 the compiler may also employ \texttt{SIGTRAP} for specific trap instructions (e.g., \texttt{brk}), but \tool{} focuses exclusively on \texttt{SIGILL} on x86\_64, where \texttt{SIGTRAP} is unrelated to CFI, see Appendix~\ref{app:isa_differences}.} caused by an \texttt{ud2} instruction.
When such a signal is received, \tool{} captures the faulting instruction pointer (\texttt{RIP}) and reconstructs the immediate calling context (e.g., return address at \texttt{[RBP + 8]}). 
To attribute the violation to high-level program entities, these addresses are resolved using \textsf{Radare2}~\cite{radare2,radare2book}, an open-source reverse-engineering and binary-analysis framework that provides disassembly, symbol resolution, and control-flow graph introspection. 
By querying \textsf{Radare2} for function boundaries, instruction metadata, and source-level mappings embedded in the DWARF debugging format information, \tool{} translates raw runtime addresses into precise code locations, such as functions, call sites, and translation units. 
Combining this static introspection with dynamic \texttt{ptrace} state enables \tool{} to pinpoint the exact control-flow edge rejected by the CFI mechanism (Section~\ref{sec:semantic_challengescfi}).

\begin{figure}[!htb]
    \centering
    \includegraphics[width=0.65\linewidth]{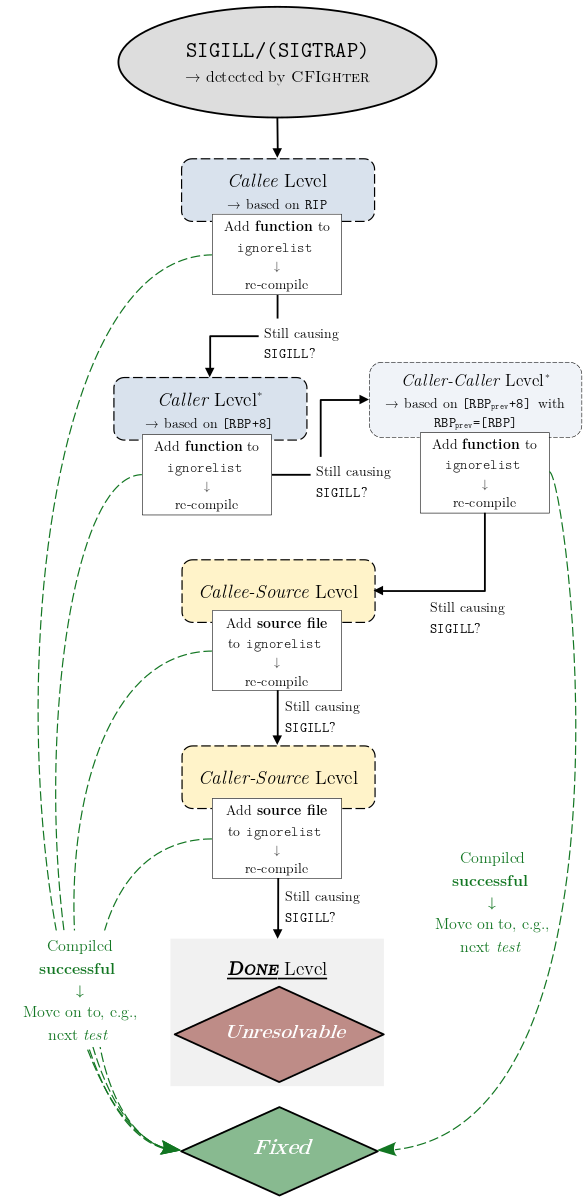}
    \caption{Escalation ladder used by \tool{} during automated fault localization.
	Each CFI violation triggers a structured progression from function-level ignores (\textit{callee} $\rightarrow$ \textit{caller}) toward translation-unit-level exceptions (source file of the \textit{callee} and then \textit{caller}), terminating either in a successful fix or an \textit{Unresolvable} classification.\\
    \textit{Illustration based on example of \texttt{x86\_64}, see Appendix~\ref{app:isa_differences}.}}
    \label{fig:escalation_levels}
\end{figure}

\tool{} searches for the smallest enforcement scope (minimal ignorelist) that eliminates the violation, following the escalation ladder in Figure~\ref {fig:escalation_levels}. 
The analysis begins at the most fine-grained scope: the faulting \textit{callee}. 
Suppose re-execution still triggers \texttt{SIGILL}. 
In that case, we broaden the exception scope to include the \textit{caller}, then the caller's caller, and finally the entire translation unit on either side of the call chain, e.g., \textit{callee}'s source file. 
Escalation halts once the violation is resolved, ensuring the ignorelist remains as small and localized as possible. 
If all escalation levels fail, the violation is classified as \textit{Unresolvable}, typically indicating an uninstrumented external library or a fundamental incompatibility between the program's semantics and the CFI policy.

By strictly separating build-time visibility adjustments from runtime policy diagnosis, and by interpreting \texttt{SIGILL} as the sole indicator of CFI violations on \texttt{x86\_64}, \tool{} achieves a robust and principled self-healing process. 
Visibility fixes restore structural correctness; CFI-policy diagnostics restore semantic compatibility. 
Together, these mechanisms enable \tool{} to instrument multi-utility, modular C/C++ projects with strong CFI guarantees while preserving functionality.

\subsubsection{Testing Phase -~\Circled[inner color=white, fill color=black]{\textnormal{\textsf{C}}}}\label{sec:testing_phase}
This phase validates the correctness and stability of all compiled variants. 
For each executable, \tool{} executes the test suites defined in the configuration file, capturing outputs, return codes, and runtime signals. 
Test results from the baseline and CFI-instrumented builds are compared to identify behavioral deviations. 
When a test fails under CFI but succeeds in the baseline, \tool{} classifies the discrepancy as either a \emph{functional non-CFI policy
violations} or a CFI policy violation. 
The former are test failures that occur only under the CFI-instrumented build but do not raise a CFI trap (e.g., \texttt{SIGILL}). 
The program executes normally, but its behavior deviates from the expected output or the test's semantics. 
Typical causes include optimization differences introduced by instrumentation, changes in evaluation order, or latent, undefined behavior exposed by additional checks. 
Although these failures do not represent CFI policy violations, they remain CFI-induced and are therefore reported separately from both baseline failures and explicit CFI traps.  
In the latter violation category, the system returns control to the analysis phase, which either updates the ignorelist or adjusts symbol visibility. 
This iterative process continues until all test cases pass or all remaining failures are classified as benign, ensuring that CFI protection does not compromise functional behavior.

\subsection{Implementation}
\tool{} is implemented in roughly 4k lines of \textit{Rust} code. 
The main benefit of using Rust is that, through its low-level control, it allows direct usage of \texttt{ptrace} for process tracing and signal interception.  
This enables \tool{} to monitor the execution of instrumented binaries, capture runtime faults such as \texttt{SIGILL}, and thus accurately classify unintended CFI policy violations. 

The framework follows an event-driven architecture comprising a build orchestrator, a runtime monitor, and an adaptive repair engine. 
The orchestrator automates baseline and CFI builds using existing toolchains and captures compiler and linker output for analysis. 
The runtime monitor supervises each verification execution and test process via \texttt{ptrace}, recording signals to identify policy violations or functional regressions. 
The results are processed by the repair engine, which applies targeted fixes such as visibility adjustments, ignorelist updates, and, upon changes, selective recompilation. 

\tool{} maintains a fully automated feedback loop until either all tests succeed or all possible CFI violations are resolved. 
The final output includes an HTML report summarizing CFI coverage across LLVM's forward-edge variants, build statistics, and violation classifications. 
We open-source \tool{}'s implementation in Section~\ref{sec:data}.

\subsection{Evaluation Setup and Subject Programs}
\label{sec:setup}
To assess the generality and robustness of \tool{}, we selected four representative GNU userland projects that vary in size, modularity, and build complexity:
\begin{enumerate}
    \item \texttt{coreutils}\footnote{\url{www.gnu.org/software/coreutils/}} (version \textit{9.6.43-47204}), 
    \item \texttt{diffutils}\footnote{\url{https://cgit.git.savannah.gnu.org/cgit/diffutils.git}} (version \textit{3.11.8-362a}), 
    \item \texttt{findutils}\footnote{\url{https://cgit.git.savannah.gnu.org/cgit/findutils.git}} (version \textit{4.10.0.42-faa1-dirty}), and
    \item \texttt{util-linux}\footnote{\url{https://github.com/util-linux/util-linux}} (version \textit{v2.42-start-826})
\end{enumerate}
These programs provide a diverse set of real-world C/C++ codebases that collectively exercise static linking, shared-library interaction, and extensive test suites, making them well-suited for evaluating automatic CFI integration.

All experiments were conducted on a Fedora~42 (x86\_64) system equipped with an Intel~Core~i7 processor and 32~GB of RAM. 
Each project was compiled using \texttt{clang}~20.1.8 with full link-time optimization (\texttt{-flto}) and hidden LTO visibility (\texttt{-fvisibility=hidden}) enabled. 
For comparison, \tool{} generated both uninstrumented \emph{baseline} builds and CFI-protected variants. 
The latter were compiled under each of LLVM's seven forward-edge CFI variants: \texttt{cfi-icall}, \texttt{cfi-vcall}, \texttt{cfi-nvcall}, \texttt{cfi-mfcall}, \texttt{cfi-derived-cast}, \texttt{cfi-unrelated-cast}, and \texttt{cfi-cast-strict}. 
All builds were executed and analyzed under \texttt{ptrace}-supervised conditions to capture signals and runtime terminations for violation classification.

Each project's existing tests were used to establish behavioral equivalence between baseline and CFI-instrumented variants. 
The framework automatically executed these test suites after every rebuild, recorded output and return codes, and compared them against baseline results. 
Any discrepancy triggered the analysis phase, which classified the fault as a visibility error, a semantic policy violation, or a non-CFI-related failure.

The first three subjects (\texttt{coreutils}, \texttt{diffutils}, and \texttt{findutils}) are multi-utility projects with comparatively simple internal architectures, minimal internal library layering, and limited reliance on dynamic linking.
Their components share small helper modules but do not rely on large, independently structured internal libraries. 
These characteristics make them suitable as initial validation cases for assessing baseline CFI compatibility and analyzing the behavior of \tool{} under well-structured control-flow patterns. 
In contrast, \texttt{util-linux} exhibits markedly higher structural diversity. 
The project combines numerous semi-independent utilities with several large internal libraries, including \texttt{libfdisk}, \texttt{libblkid}, \texttt{libmount}, \texttt{libuuid}, and \texttt{libsmartcols}, and contains more than $6K$ compiled functions spread across heterogeneous subsystems. 
Its combination of table-driven parsers, callback-based infrastructure, state-machine dispatch logic, and compiler-generated jump tables (JT) creates a highly diverse set of control-flow graphs. 
For this reason, \texttt{util-linux} serves as a representative stress test for evaluating \tool{}'s ability to operate in complex modular architectures and to automatically repair forward-edge CFI incompatibilities. 
The combined dataset thus spans both compact, monolithic utilities and modular, multi-component systems, providing a realistic cross-section of open-source C/C++ software to evaluate \tool{}'s applicability across different software structures.

\section{Results}
This section reports the behavior of \tool{} on four real-world GNU projects described in Section~\ref{sec:setup}.  
We begin by characterizing the static control-flow space that forward-edge CFI must protect. 
We then evaluate the degree to which the existing test suites exercise that target set. 
Finally, we analyze the CFI-induced failures produced during testing and assess the extent to which \tool{} can automatically restore compatibility. 
The goal is to understand both the inherent difficulty of applying strict forward-edge CFI to widely deployed system utilities and the practical effectiveness of our compatibility-recovery mechanism.

\subsection{Subject Systems and Static Size}\label{sec:results:subjects}

Table~\ref{tab:results-size} summarizes the size of the instrumented codebase for each subject project. 
Even relatively small utilities expose a large number of indirect control-flow sites. 
\begin{table}[!htb]
  \centering
    \caption{Subject systems and static size of the CFI-instrumented codebase.}
    \label{tab:results-size}
  \begin{adjustbox}{width=0.75\linewidth, center}
  \begin{tabular}{lrrrr}
    \toprule
    Project & LoC & \#Functions & IR LoC & \#Indirect call sites \\
    \midrule
    \rowcolor{light-gray}
    \texttt{coreutils}  & 42\,154 & 4\,273 & 566\,559 & 16\,954 \\
    \texttt{diffutils}  & 5\,992  & 175  & 111\,935 & 3\,440 \\
    \rowcolor{light-gray}
    \texttt{findutils}  & 6\,359  & 315  & 147\,010  & 4\,942  \\
    \texttt{util-linux} & 29\,960 & 6\,455 & 1\,066\,509 & 28\,242 \\
    \bottomrule
  \end{tabular}
  \end{adjustbox}
\end{table}
For instance, \texttt{diffutils} contains fewer than $6K$ lines of code, yet includes more than $3\,400$ indirect transfers. 
At the other end of the spectrum, \texttt{util-linux} contains nearly $30K$ such sites. 
Since forward-edge CFI must classify and monitor every indirect transfer, these numbers illustrate the practical difficulty of enforcing strict CFI in real-world C code bases.

In order to understand the structure of the indirect control flow that \tool{} must handle, we derive a more detailed breakdown using the LLVM IR emitted by Clang. 
Table~\ref{tab:results-cf-breakdown} distinguishes between key categories, including function-pointer (FP) calls, callbacks, virtual dispatch, and compiler-generated jump tables (JT). 
\begin{table*}[!t]
  \centering
  \caption{LLVM IR control-flow categories per project.}
  \label{tab:results-cf-breakdown}
    \begin{adjustbox}{width=\linewidth, center}
      \begin{tabular}{lrrrrrr}
        \toprule
        Project & FP calls & C++/virtual calls & Callback stores & JT (switch) & JT (lowered) & Inline ASM \\
        \midrule
        \rowcolor{light-gray}
        \texttt{coreutils}  & 338 & 6 & 2\,316 & 776 & 13\,448 & 70 \\
        \texttt{diffutils}  & 46   & 3   & 318 & 200 & 2\,873  & 0  \\
        \rowcolor{light-gray}
        \texttt{findutils}  & 63   & 5   & 380 & 253 & 4241  & 0  \\
        \texttt{util-linux} & 191 & 20 & 1\,461 & 1\,458 & 25\,056 & 56 \\
        \bottomrule
      \end{tabular}
  \end{adjustbox}
\end{table*}
The last category dominates across all projects. 
Two distinct structures appear in the IR. 
High-level jump tables correspond to \texttt{switch} instructions that remain structured in the IR, which allows CFI to reason directly about the control-flow alternatives. 
Lowered jump tables arise when optimizations translate the original multi-way branch into a low-level indexed dispatch by materializing a table of code pointers or block addresses and emitting a computed indirect branch. 
The lowered form provides little structural information and behaves similarly to unstructured indirect control flow, which increases the burden on type-based CFI. 
Appendix~\ref{app:lowered-jt} provides additional explanation and examples on jump tables. 

Both \texttt{coreutils} and \texttt{util-linux} contain exceptionally large numbers of lowered jump tables, with 13\,448 and 25\,056 occurrences respectively. 
Function pointers and callbacks appear in moderate numbers in all projects, while virtual dispatch is rare because the code bases are almost entirely written in C. 
Inline assembly is present in \texttt{coreutils} and \texttt{util-linux}. 
Although infrequent, inline assembly nevertheless introduces regions where the CFI mechanism cannot rely on type-level metadata and thus cannot be applied. 

Taken together, the static analysis shows that \tool{} must handle tens of thousands of indirect control-flow sites, many of which originate from compiler transformations rather than explicit developer intent. 
A smaller but non-negligible portion of the sites arises from inline assembly and loosely typed callback mechanisms. 
These factors make the evaluation in this section representative of the challenges that arise when deploying CFI in large, mature system utilities.

\subsection{Test-Suite Coverage of CFI-Protected Code}\label{sec:results:coverage}
We next assess how thoroughly the individual test suites exercise the CFI-protected code.
Table~\ref{tab:results-coverage} reports the line and function coverage observed under the same compiler configuration used for \tool{}.

\texttt{Coreutils} benefits from an extensive upstream test suite and achieves more than 75\% line coverage and over 80\% function coverage, providing strong empirical support for the compatibility results reported in Section \ref{sec:results:failures}. 
In contrast, the remaining projects achieve between 40–60\% line coverage, leaving larger untested regions. 
\begin{table}[!htb]
  \centering
  \caption{Test-suite coverage of the CFI-instrumented code base, excluding \textit{gnulib-tests}.}
  \label{tab:results-coverage}
      \begin{adjustbox}{width=0.65\linewidth, center}
  \begin{tabular}{lrr}
    \toprule
    Project             & Line coverage     & Function coverage \\
    \midrule
    \rowcolor{light-gray}
    \texttt{coreutils}  &       75.3\%      &      81.6\%      \\
    \texttt{diffutils}  &       49.9\%      &      53.4\%      \\
    \rowcolor{light-gray}
    \texttt{findutils}  &       42.3\%      &      53.0\%      \\
    \texttt{util-linux} &       49.4\%      &       62.4\%      \\
    \bottomrule
  \end{tabular}
  \end{adjustbox}
\end{table}
Since \tool{} operates only on executed code paths, the dynamic results should still be interpreted as lower bounds on the number of latent CFI issues, especially for projects with more modest coverage.

\subsection{CFI-Induced Failures and Automatic Recovery}\label{sec:results:failures}
Executing the CFI-enabled binaries under their respective test suites results in a clear separation among the subjects. 
\texttt{coreutils}, \texttt{diffutils}, and \texttt{findutils} complete their entire tests without any CFI-induced failures. 
The corresponding \tool{} reports contain no violation tables and list no tests that terminated due to invalid indirect control-flow transfers. 
Given the limited coverage in some of these projects, the absence of failures indicates compatibility along the executed subset, but does not imply full compatibility across the entire code base.

In contrast, \texttt{util-linux} exhibits a far more heterogeneous control-flow structure.
This project contains several large libraries, such as \texttt{libmount}, \texttt{libsmartcols}, and \texttt{libfdisk}, and its control-flow topology is considerably more heterogeneous. 
During testing, \texttt{util-linux} produces CFI violation events (\texttt{SIGILL}) across 48 individual tests. 
The failures originate from six distinct code regions, including two files in \texttt{libfdisk}, one in \texttt{libmount}, and two in \texttt{libsmartcols}. 
These sections of the code base rely on table-driven parsing, callback-dependent formatting logic, and state-machine style dispatch patterns that are difficult for type-based CFI to represent precisely. 

Despite this complexity, 46 violations that match known escalation patterns are automatically resolved by \tool{}. 
After each violation is observed, \tool{} applies its iterative escalation strategy, re-runs the affected test, and validates that the repaired configuration no longer triggers a CFI exception. 
Only two violations remain unresolved at the end of the process. 
Both occur in deeply optimized parsing and formatting paths, where compiler transformations obscure the original control-flow structure. 

Taken together, these findings show that strict CFI is not inherently compatible with large, multi-component utilities such as \texttt{util-linux}. 
At the same time, the results indicate that \textsc{CFI-ghter} can repair the vast majority of practical incompatibilities in a fully automated manner, without disabling CFI across entire libraries or requiring developer intervention.

\begin{table*}[!t]
  \centering
  \caption{Overall Test-based Results of \tool{}}
  \label{tab:results-tests}
    \begin{adjustbox}{width=\linewidth, center}
      \begin{tabular}{lR{1cm}R{1.5cm}R{1cm}R{1.5cm}R{1.5cm}R{1.5cm}R{1.5cm}}
        \toprule
        Project                         & 
        Total Tests                     & 
        CFI-Policy Violations           &  
        Fixed Tests                     & 
        Unresolved Tests                & 
        Non-CFI Policy Violations       & 
        Total CFI Coverage (IR-based)   &
        Duration (\textit{hh:mm:ss})    \\
        \midrule
        \rowcolor{light-gray}
        \texttt{coreutils}              &      
            526                         &       
              0                         &        
              0                         &       
              0                         &  
              0                         & 
            \textcolor{ao}{100\%}       &  
            00:27:33                    \\
            
        \texttt{diffutils}              &      
            339                         &       
              0                         &        
              0                         &       
              0                         &  
              0                         &   
            \textcolor{ao}{100\%}       &        
            00:10:57                    \\
            
        \rowcolor{light-gray}
        \texttt{findutils}              &       
             25                         &       
              0                         &        
              0                         &       
              0                         &  
              0                         & 
            \textcolor{ao}{100\%}       &  
            00:13:13                    \\
            
        \texttt{util-linux}                     &      
            234                                 &      
             48                                 & 
            \textcolor{ao}{\textbf{46}}         & 
            \textcolor{rosevale}{\textbf{2}}    &     
            \textcolor{neoncarrot}{\textbf{4}}  & 
            \textcolor{ao!60}{89.46\%}        &
            02:15:50                            \\
        \bottomrule
      \end{tabular}
    \end{adjustbox}
\end{table*}
The overall duration of \tool{}'s analysis reflects not only the size of each project and the cost of its test suite, but also the number of visibility adjustments required during the repair process. 
Importantly, this cost is incurred only during the first application of \tool{} to a given codebase. 
Once the visibility configuration has converged and the project has been repaired, subsequent runs, whether for testing, reanalysis, or when applying \tool{} after unrelated code modifications, can reuse the established configured source code. 
As a result, re-running \tool{} does not repeat the full sequence of visibility adjustments, significantly reducing the analysis time compared to the initial run. 
For instance, the extended execution time observed for \texttt{util-linux} stems from the numerous visibility adjustments applied during the first repair cycle. 
Once these adjustments are in place, subsequent applications of \tool{} complete dramatically faster, reducing the total runtime from 02:15:50 to  01:26:03.

\subsection{Case Study: \texttt{util-linux}}\label{sec:case_study}
All observed violations were raised by Clang's \texttt{cfi-icall} mechanism; any other CFI policy caused no failures. 
The remaining projects (\texttt{coreutils}, \texttt{diffutils}, and \texttt{findutils}) exhibited no CFI violations at all. 
A closer examination of the \texttt{util-linux} codebase helps contextualize the observed violations. 
The project comprises more than 6\,000 compiled functions and nearly 30\,000 indirect control-flow sites. 
Its internal libraries (\texttt{libblkid}, \texttt{libmount}, \texttt{libsmartcols}, \texttt{libfdisk}, and \texttt{libuuid}) combine heterogeneous components, including filesystem probing, mount-table parsing, terminal formatting, column layout, and partition manipulation. 
As a result, multiple dispatch styles coexist within the same binary, including table-driven parsers, callback-based formatting pipelines, and state-machine dispatch functions. 
These patterns interact in non-trivial ways with type-based CFI. 
We avoid qualitative labels such as ``complex'' and instead characterize \texttt{util-linux} by its objective structural properties: 
a multi-library architecture, numerous heterogeneous dispatch mechanisms, and the largest indirect control-flow space among our subjects.

During testing, \texttt{util-linux} triggers CFI violation events across 48 individual tests, all of them based on \texttt{cfi-icall}. 
These violations are highly clustered. 
Thirty violations originate from \texttt{libsmartcols/src/filter-param.c}, a translation unit implementing callback-driven formatting and filtering operations with several indirect dispatch sites. 
A further eleven violations arise in \texttt{libfdisk/src/ask.c}, which implements interactive and script-driven state transitions in the \texttt{fdisk} utility. 
Four violations occur in \texttt{libmount/src/tab.c}, which uses a compact table-driven parsing logic that relies on type-heterogeneous handler structures. 
A single additional violation originates in \texttt{libfdisk/src/script.c}, exercised by the \texttt{libfdisk/gpt} test. 
No CFI violations originate from \texttt{libsmartcols/src/print.c}: 
although the \texttt{fincore/count} test fails there, this behavior constitutes a functional non-CFI policy violation in the sense defined in Section~\ref{sec:testing_phase}, and does not reflect a CFI policy violation.

Execution traces and dynamic linking information show that several failing \texttt{lsblk} and \texttt{column} tests hit the same translation units responsible for the observed violations. 
For instance, the \texttt{column/table} test reaches dispatch logic in \texttt{libsmartcols/src/line.c} through the shared library \texttt{libsmartcols.so.1.1.0}. 
In contrast, the \texttt{lsblk/lsblk} test does \textit{not} trigger a CFI failure in  \texttt{libsmartcols/src/print.c}; its observed behavior is unrelated to CFI enforcement. 
These relationships show how distinct high-level utilities converge on shared dispatch and formatting routines, making it unsurprising that violations concentrate in a small number of translation units.
A broader perspective confirms that these ignorelist entries are highly localized. 
If the entire \texttt{libsmartcols} directory is excluded, effectively disabling CFI for that library as a whole, only a single test (\texttt{lsfd/column-xmode}) fails. 
That failure is purely functional and also occurs in the baseline tests, indicating it is not CFI-related. 
Under such a coarse-grained suppression strategy, all four problematic behaviors observed in the fine-grained evaluation (the two unresolvable CFI violations and the two functional failures) disappear. 
However, this comes at a substantial cost to enforcement strength: only 84.27\% of indirect call sites remain protected in this configuration, compared to the significantly higher protection levels achieved by \tool{} 's targeted, escalation-based repair. 
The remaining \texttt{lsfd/column-xmode} failure is still triggered under every CFI variant we evaluated and disappears only when the project is compiled entirely without CFI. 
This indicates that the issue is not specific to \tool{}'s repair strategy but reflects an inherent interaction between CFI instrumentation and the tested code path in \texttt{libsmartcols}.

The escalation strategy implemented by \tool{} successfully repairs nearly all observed violations. 
Most cases are resolved by switching from per-call-site to per-function enforcement, while several more complex regions require escalation to the translation-unit level. 
These adjustments affect only the local region that contains atypical dispatch patterns, preserving strict enforcement across the vast majority of the code base. 
Two violations remain unresolved. 
Both occur in translation units that combine multi-stage dispatch with aggressive compiler transformations. 
Since the prototype does not record faulting program counters, we conservatively refrain from attributing these failures to specific compiler transformations and treat them as opportunities to refine the escalation mechanism in the future.

The \textit{visibility}-repair process for \texttt{util-linux} requires 57 iterations in total: 
46 during the initial build phase and an additional 11 during the subsequent testing phase. 
Multiple iterations are necessary because \tool{} relies on compiler and linker feedback to identify the specific call sites involved in each violation. 
During the initial build phase, only violations that appear in the compiler's output stream are visible, and that output is subject to length limitations that may suppress additional entries. 
Additional violations emerge only when the affected binaries are executed, as they stem from code paths reachable only after specific objects are linked or after later-discovered call sites are exercised by the test suite. 
Consequently, the iterative approach is inherent to the structure of the build and testing pipeline: each repaired violation reveals additional call sites and dispatch patterns that were not previously observable, necessitating further refinement of the visibility configuration until a fixed point is reached.

\begin{table}[!htb]
  \centering
  \small
  \caption{Distribution of CFI violations across \texttt{util-linux} translation units.}
  \label{tab:util-linux-violations}
        \begin{adjustbox}{width=\linewidth, center}
  \begin{tabular}{lrr}
    \toprule
    Source File / Library & Violations & Triggering Tests \\
    \midrule
    \rowcolor{light-gray}
    \begin{tabular}[c]{@{}l@{}}\texttt{libsmartcols/}\texttt{src/}\\\texttt{filter-param.c}\end{tabular} & 30 & \texttt{lsfd/*}, \texttt{column/*}, \texttt{findmnt/*}, \dots \\
    \texttt{libfdisk/src/ask.c}             & 11 & \texttt{fdisk/*} \\
        \rowcolor{light-gray}
    \texttt{libmount/src/tab.c}             & 4  & \texttt{findmnt/*} \\
    \begin{tabular}[c]{@{}l@{}}\texttt{libfdisk/src/}\\\texttt{script.c}\end{tabular}          & 1  & \begin{tabular}[c]{@{}r@{}}\texttt{libfdisk/gpt}\\~\end{tabular} \\
    \bottomrule
  \end{tabular}
  \end{adjustbox}
\end{table}

\begin{figure}[!htb]
\centering

\begin{minipage}{0.45\linewidth}
\centering
\begin{adjustbox}{width=\linewidth, center}
\begin{tikzpicture}
\pie[
    color={olivine, neoncarrot, rosevale},
    before number=\bfseries,
    after number=\%,
    radius=2.5,
    explode=0.2
]{
    86.29/,
    10.54/,
    3.17/
}
\end{tikzpicture}
\end{adjustbox}
\subcaption{Enforcement distribution aggregated per function}
\label{fig:pie-a}
\end{minipage}
\hfill
\begin{minipage}{0.45\linewidth}
\centering
\begin{adjustbox}{width=\linewidth, center}
\begin{tikzpicture}
\pie[
    color={olivine, neoncarrot, rosevale},
    before number=\bfseries,
    after number=\%,
    radius=2.5,
    explode=0.2
]{
    89.46/,
    7.53/,
    3.01/
}
\end{tikzpicture}
\end{adjustbox}

\subcaption{Enforcement distribution aggregated per indirect call site}
\label{fig:pie-b}
\end{minipage}
\begin{adjustbox}{width=0.6\linewidth, center}
    \begin{tabular}{ll}
        &\\
        \textcolor{olivine}{$\blacksquare$} & Protected $\Rightarrow$ \textit{CFI-enforced} \\
        \textcolor{neoncarrot}{$\blacksquare$} & \texttt{"default"}-visibility $\Rightarrow$ \textit{unprotected} \\
        \textcolor{rosevale}{$\blacksquare$} & Ignored (escalation-level exclusion) $\Rightarrow$ \textit{unprotected} \\
    \end{tabular}
\end{adjustbox}
\caption{
    Final CFI enforcement distribution for \texttt{util-linux} after \tool{}’s iterative repair process.
    Subfigure~\subref{fig:pie-a} aggregates enforcement at the function level, while  
    Subfigure~\subref{fig:pie-b} aggregates enforcement across all indirect call sites.  
    Together, they illustrate that strict CFI remains active across the dominant share of the code base, with only small, localized regions requiring relaxation.
}
\label{fig:results}
\end{figure}
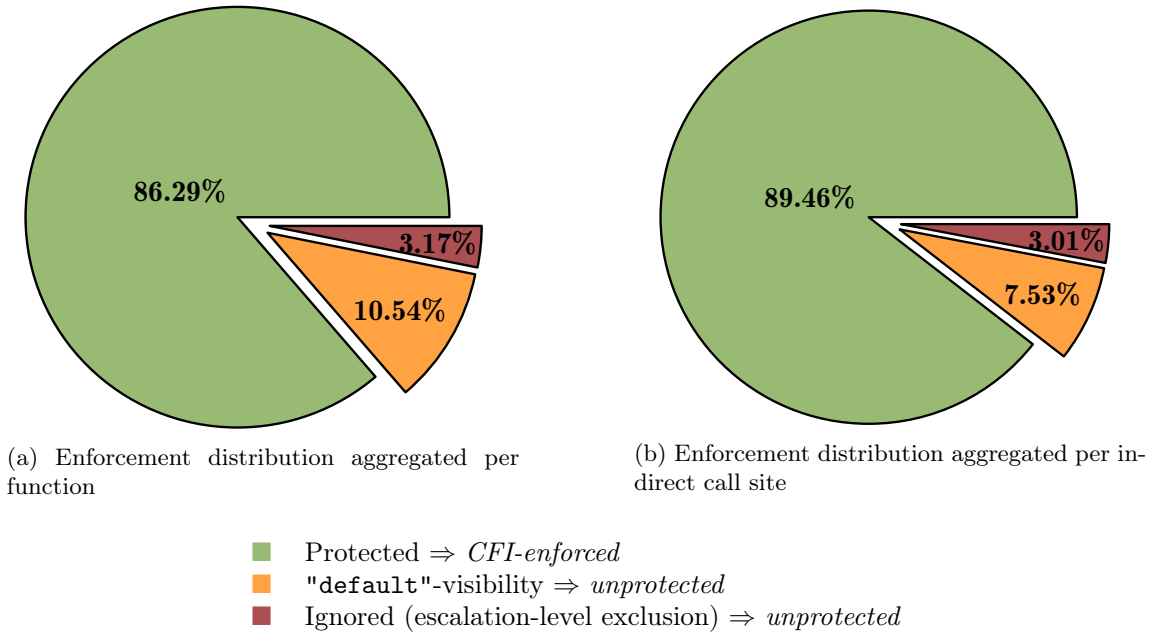

Figure~\ref{fig:results}\subref{fig:pie-a} and Figure~\ref{fig:results}\subref{fig:pie-b} summarize the final enforcement distribution that \tool{} applies to \texttt{util-linux}. 
The first view (Figure~\ref{fig:results}\subref{fig:pie-a}) reports the fraction of protected, default-visibility, and ignored items when aggregated at the function level, whereas the second view (Figure~\ref{fig:results}\subref{fig:pie-b}) reflects the same distribution across all indirect call sites (Table~\ref{tab:results-size}). 
While the two perspectives differ in magnitude, both confirm that strict CFI remains active across the dominant fraction of the code base, with only a comparatively small portion requiring relaxation.

All entries in the resulting ignorelist originate from escalation at the callee–source level. 
In other words, every ignored item corresponds to a translation unit whose internal dispatch patterns or compiler transformations prevented \tool{} from maintaining a safe per-call-site or per-function policy. 

\subsubsection*{Manual Inspection Summary}
To contextualize the distribution of violations, we manually inspected the affected tests and their associated execution paths. 
For each violation reported by \tool{}, we identified the test that triggered it, located the invoked binary, and examined the internal libraries and translation units used during execution. 
This analysis required no modifications to \tool{} itself: 
all information was derived from dynamic linking data, test-suite output, and source review.

Two findings emerged. First, a significant fraction of violations arises in shared formatting and dispatch code, especially within \texttt{libsmartcols}, explaining why distinct tests (e.g., \texttt{lsfd}, \texttt{column}) surface failures in the same few translation units. 
Second, the numerical clustering in Table~\ref{tab:util-linux-violations} reflects shared internal pathways rather than multiple independent defects: 
translation units such as \texttt{libsmartcols/src/filter-param.c} and \texttt{libfdisk/src/ask.c} are executed by many tests. 
In contrast, those with few violations correspond to a more narrowly exercised code. 
Although the prototype does not emit precise faulting addresses, this mapping reliably identifies which components are responsible for the observed incompatibilities and how they are reached during testing.

Overall, the results demonstrate both the scale and the intricacy of enforcing strict forward-edge CFI in mature GNU utilities. 
Across all subjects, \tool{} maintains strong protection while accommodating diverse dispatch patterns, compiler-generated control-flow transformations, and the practical limitations of incomplete test coverage. 
The evaluation further shows that most incompatibilities are highly localized, that the vast majority of indirect control-flow transfers remain fully protected, and that \tool{}'s iterative escalation strategy is sufficient to restore compatibility even in large, heterogeneous code bases such as \texttt{util-linux}. 
These findings indicate that precise, automated configurability is not only feasible but also essential for deploying strong CFI in real-world systems software.

Taken together, the test-suite coverage results (Section~\ref{sec:results:coverage}) provide important context for interpreting these outcomes. 
Three of the four subject projects (\texttt{coreutils}, \texttt{diffutils}, and \texttt{findutils}) exhibit \emph{100\%} CFI-policy coverage along all executed paths, supported in the case of \texttt{coreutils} by an extensive upstream test suite that exercises a substantial portion of the code base. 
For the remaining project, \texttt{util-linux}, test coverage between roughly 40--60\% implies that \tool{} achieves robust CFI integration along the exercised paths, while untested regions may still contain latent incompatibilities.

\section{Discussion}
\subsection{Implementation Decisions}\label{sec:implementation_decisions}
We deliberately chose to implement \tool{} on top of \texttt{ptrace} rather than a debugger, since the latter can aggravate CFI fault-causation isolation.  
When a CFI violation occurs, the compiler emits a \texttt{ud2} instruction that raises \texttt{SIGILL} and transfers control into a short failure handler. 
Debuggers such as GDB or LLDB catch the resulting signal only after this trampoline executes, at which point the register and stack state have already been modified to enter the runtime handler. 
The technical details are explained in more depth in Appendix~\ref{app:ptrace-unwinding}. 
As a result, the backtrace often attributes the fault to the sanitizer runtime rather than the original call site, hiding the true faulting instruction. 
In contrast, \texttt{ptrace} receives the stop directly from the kernel at the moment of the \textit{Illegal instruction}, exposing the raw register file and program counter before any unwinding or handler invocation. 
This low-level access allows \tool{} to identify the exact instruction and call target responsible for the violation, something that higher-level debuggers cannot reliably reconstruct. 

A potential extension of this work could involve supporting cross-DSO (\textit{dynamically shared objects}) CFI, which protects control-flow transfers across dynamically linked components.  
However, its reliance on runtime registration and per-library compilation with consistent CFI metadata poses significant practical challenges for automated frameworks such as \tool{}.  
Generic integration would require rebuilding all dependencies with identical compiler versions and visibility policies while tolerating nondeterministic runtime metadata registration.  
These conditions are rarely met in heterogeneous software stacks, meaning ``\textit{real-world, diverse codebases}", not perfectly controlled, uniform build environments. 
This makes cross-DSO CFI practically infeasible for large-scale, project-agnostic automation without sacrificing reproducibility and enforcement precision.

\subsection{Implementation Scope and Constraints}\label{sec:limitations}
While \tool{} substantially automates the adaptation of large C/C++ projects to LLVM's forward-edge Control-Flow Integrity (CFI), its implementation necessarily reflects a number of practical constraints arising from compiler toolchains, system interfaces, and the design trade-offs described in Section~\ref{sec:design}. 
These constraints do not diminish the utility of \tool{}, but they delineate the conditions under which the framework delivers reproducible, precise results.

\subsubsection{Focus on Intra-DSO CFI}
As discussed in Section~\ref{sec:implementation_decisions}, \tool{} targets the standard \emph{intra-DSO} model used by Clang/LLVM's forward-edge CFI~\cite{tice2014enforcing,LLVMdoc,androidcfi}. 
This model assumes that all code participating in protected control-flow transfers is available at compile and link time, for example, in statically linked binaries or single-DSO builds, so that LLVM can analyze and instrument all indirect call targets consistently. 
In practice, this requires that the project and its dependencies are compiled under the same CFI configuration, ensuring that the compiler has complete visibility into the set of legitimate call targets. 
Supporting cross-DSO CFI would require rebuilding every dependent library with identical compiler versions, LTO modes, and CFI metadata formats, and then registering type information deterministically at program start. 
Such tight coordination is rarely achievable in heterogeneous software ecosystems, which often combine system libraries, distribution packages, and third-party modules compiled under different toolchains. 
For this reason, cross-DSO CFI, which has been shown to be fragile even in controlled environments~\cite{burow2017control}, remains beyond the practical scope of automated frameworks like \tool{}.

\subsubsection{Architecture-Specific Trap Semantics}
\tool{}'s runtime diagnosis relies on the trap behavior of LLVM's forward-edge CFI on \texttt{x86\_64}, where failed checks expand to a \texttt{ud2} instruction that reliably delivers a \texttt{SIGILL} before any runtime handler executes. 
This predictable signaling behavior allows \tool{} to intercept violations via \texttt{ptrace}, recover the faulting instruction pointer, and reconstruct the local calling context.

Other architectures exhibit different trap and unwinding semantics. 
AArch64, for example, may encode CFI traps using \texttt{brk} instructions and employs a calling convention based on link-register returns. 
Depending on the compilation settings, the compiler may also emit pointer-authentication instructions (PAC), and stack unwinding is described through DWARF CFI metadata rather than the \texttt{RBP}-chained frames commonly found on \texttt{x86\_64}. 
These differences mean that the procedure used to recover the faulting state, including register interpretation, return-address reconstruction, and stack frame enumeration, cannot be directly ported.

Supporting such platforms would therefore require architecture-specific \texttt{ptrace} unwinding routines and tailored signal-handling logic (see Appendix~\ref{app:ptrace-unwinding}). 
For this reason, the current prototype focuses on the \texttt{x86\_64} intra-DSO CFI model, where trap semantics and unwinding behavior are uniform, stable, and readily amenable to low-level instrumentation.

\subsubsection{Dependence on Symbol and Debug Information}
Accurate attribution of CFI violations requires resolving the faulting instruction pointer to the corresponding function, call site, or translation unit. 
This process, commonly known as \textit{symbol resolution}, relies on DWARF debug information and the symbol tables embedded in unstripped ELF binaries. 
These metadata describe function boundaries, inlining structure, and source locations, enabling precise identification of the calling context.
When binaries are stripped, however, symbol and debug sections are removed. 
In such cases, only raw code addresses remain, and neither function names nor source-level mappings can be recovered directly. 
As a result, CFI violations can be attributed only to coarse-grained address ranges, preventing \tool{} from deriving fine-grained ignorelist entries such as individual call sites or specific callee functions.

\tool{} mitigates this loss of information by invoking \textsf{Radare2}~\cite{radare2} on the target binary. 
When symbolic metadata is absent, \textsf{Radare2} falls back to disassembly-based heuristics that infer approximate function boundaries from instruction patterns and control-flow structure. 
While this allows the analysis to proceed, the lack of symbolic metadata reduces the precision of the repair process and may force \tool{} to escalate to broader enforcement scopes than would be necessary in fully annotated builds.

 \subsubsection{Locality of Violation Resolution}
As described in Section~\ref{sec:analysis_phase}, \tool{} uses a structured escalation ladder (Figure~\ref{fig:escalation_levels}) that expands exception scope only along the observed dynamic call chain. 
This locality keeps exceptions small, deterministic, and easily auditable, but it also means that code not executed during the test suite will remain unexamined. 
This inherent dependency on dynamic coverage mirrors limitations of other dynamic instrumentation and testing systems~\cite{zeller2002simplifying,fuzzing2012sage,metzman2021fuzzbench}: rare, optional, or non-default code paths may still trigger CFI checks if executed outside the test harness.

 \subsubsection{Limited Semantically Rich Root-Cause Analysis}
\tool{} treats violations uniformly as mismatches between observed control-flow behavior and compiler-generated metadata~\cite{abadi2009control,houy2025sok}. 
The framework does not attempt to infer higher-level design idioms such as plugin registries, polymorphic callback tables, or interpreters with intentionally flexible dispatch mechanisms. 
While this keeps the repair model simple and fully automated, richer semantic reasoning could potentially reduce over-approximation in codebases with complex dynamic interfaces. 

 \subsubsection{Scope of Automated Visibility Repair}
During the Building Phase (Section~\ref{sec:building_phase}), \tool{} performs conservative automatic visibility adjustments in response to unresolved symbols caused by LTO and hidden visibility policies (Section~\ref{sec:lto}). 
These repairs are intentionally minimalistic: only the specific symbols required to restore linkability are widened. 
Projects that use custom build scripts, linker scripts, or nonstandard symbol export mechanisms may require project-specific extensions beyond the current repair model.

 \subsubsection{Limited Support for Runtime Code Generation}
Forward-edge CFI relies on compile-time type metadata and static instrumentation. 
Consequently, \tool{} is unsuitable for workloads that generate executable code at runtime, including JIT (\textit{Just-In-Time}) compilers~\cite{ducasse2025war,bauman2023renewable,niu2014rockjit,v8}, eBPF\footnote{\url{https://ebpf.io/what-is-ebpf/}} loader pipelines~\cite{sharaf2022extended}, and dynamic code patching engines, whose control-flow targets are not visible to the compiler and therefore cannot be safely instrumented~\cite{snow2013just}. 
Supporting such workloads would require integrating dynamic recompilation or runtime verification, which is beyond the scope of this work.

 \subsubsection{Forward-Edge Focus}
\tool{} currently addresses only forward-edge CFI. 
Backward-edge protections such as shadow stacks, Intel CET, or compiler-based return protection~\cite{CET, PAC} operate through separate runtime mechanisms and can interact with \texttt{ptrace} in architecture-specific ways. 
Their integration would require a significant redesign of both the trap-handling and exception-generation pipeline and is thus considered orthogonal future work.

 \subsubsection{Dependence on the Test Suite}
Finally, \tool{}’s repair process is bounded by the behavioral coverage of the project's test suite. 
Only violations exercised during the Testing Phase (Section~\ref{sec:testing_phase}) can be observed and corrected. 
Although this limitation is standard in dynamic analysis, it is particularly relevant for CFI: infrequently executed dispatch paths may still produce enforcement failures outside the automated workflow. 
Improving coverage remains an orthogonal but beneficial step toward strengthening \tool{} 's guarantees.

\section{Related Work}
Control-flow integrity (CFI) has been studied extensively across compiler instrumentation, binary rewriting, runtime monitoring, and hardware extensions. 
Foundational work by Abadi et al.\ introduced the CFI model~\cite{abadi2009control}, and subsequent surveys analyze the precision, cost, and practical limitations of existing defenses~\cite{burow2017control,houy2025sok,becker2024sok}. 
While the literature has optimized enforcement strength for more than a decade, relatively little research addresses the \textit{deployment} challenges that arise when enabling strict CFI in large, modular C/C++ systems. 
\tool{} targets precisely this gap.

\subsection{Compiler-Based CFI}
Modern compiler-integrated mechanisms such as LLVM's forward-edge CFI~\cite{tice2014enforcing,LLVMdoc,LLVMDesigndoc} represent the dominant form of fine-grained, type-based enforcement in practice. 
Major ecosystems rely on variants of this design: Chromium and Android deploy platform-specific adaptations of LLVM's CFI, whereas Firefox has long planned but not yet succeeded in enabling it due to compatibility constraints~\cite{chromiumcfi,firefoxcfi,androidcfi}. 
Microsoft Windows instead ships its own independently developed forward-edge mechanism, Control Flow Guard (CFG)~\cite{windowscfi}. 
GCC provides a complementary C++ mechanism, VTV (virtual-table verification), that validates \texttt{vtable} targets at runtime to secure virtual dispatches~\cite{vtv,vtvProposal,vtvGuide}. 
However, VTV protects only C++ virtual calls and does not constrain general function-pointer dispatch, making it a partial, rather than full, forward-edge CFI mechanism.

Research systems such as MCFI~\cite{niu2014modular} introduced modular, per-module enforcement techniques, and Per-Input CFI ($\pi$-CFI) refined target sets further by specializing the allowed callees for each concrete program input~\cite{niu2015per}. 
Although influential in exploring design space, these prototypes are not deployed in mainstream toolchains.

Empirical studies show that type-based CFI remains fragile in the presence of inconsistent prototypes, symbol-visibility mismatches, inline assembly, and
dynamic loading~\cite{farkhani2018effectiveness,muntean2019analyzing,xu2019confirm}. 
Visibility rules and cross-DSO boundaries are frequent causes of breakage in real deployments~\cite{LLVMlto,redhatVisibility,crossDSO}. 
Tools such as CCured~\cite{necula2002ccured}, CUP~\cite{burow2018cup}, and TypeSan~\cite{haller2016typesan} strengthen type correctness but do not integrate with compiler CFI nor resolve CFI-induced build or runtime errors. 
Consequently, while compiler-based CFI mechanisms are mature, their reliable deployment in complex software remains challenging.

\subsection{CFI Deployment, Compatibility, and Automated Adaptation}
A growing body of research examines the compatibility and maintainability barriers to deploying CFI at scale. NesCheck~\cite{midi2017memory}, SafeDispatch
\cite{jang2014safedispatch}, ClangMR~\cite{wright2013large}, interface sanitizers~\cite{ming2025digital}, and ABI checkers detect inconsistencies but do not consider CFI's additional visibility and type requirements, nor provide automated repair. 
Compatibility evaluations such as CONFIRM~\cite{xu2019confirm} and analyses of LLVM-CFI~\cite{muntean2019analyzing} quantify the brittleness of compiler-based
CFI, and long-term adoption studies~\cite{houy2025twenty,becker2024sok} highlight persistent engineering obstacles. 
Most closely related, Houy et al. document the substantial manual work required to deploy CFI in OpenJDK's JVM~\cite{houy2024lessons}. 
They identify type inconsistencies and dynamic-loading behavior as the dominant causes of CFI failures. In our setting, we additionally observe that symbol-visibility mismatches and weak/hidden symbol interactions can trigger similar problems. 
These studies characterize the problem but do not automate its resolution.
To our knowledge, \tool{} is the first framework to automatically \textit{detect}, \textit{classify}, and \textit{repair} build-time and runtime failures caused by strict forward-edge CFI.

\subsection{Binary, Dynamic, and Hardware Approaches}
Binary-level systems such as RetroWrite~\cite{dinesh2020retrowrite}, per-input CFI~\cite{niu2015per}, and TypeArmor~\cite{van2016tough} retrofit enforcement onto commodity binaries using rewriting or dynamic instrumentation, but lack compiler-level type information and often lose precision. 
Related COTS protections~\cite{zhang2015control} improve robustness but incur overhead.

Hardware-supported CFI mechanisms such as Intel CET~\cite{CET}, ARM Pointer Authentication~\cite{PAC,applecfi}, authenticated call stacks~\cite{liljestrand2019authenticated}, and shadow-stack systems~\cite{burow2019sok,LLVMscs} provide strong architectural protection for control-flow transfers. 
However, software cannot assume that such hardware features are present or enabled across all deployment environments, particularly when targeting heterogeneous systems, portable distributions, or legacy installations. 
As a consequence, practical deployments typically rely on compiler-based forward-edge CFI, which is widely available but sensitive to semantic mismatches between application-level dispatch patterns and the compiler's type-based policy. 
\tool{} addresses this practical compatibility boundary by repairing the cases where strict compiler-enforced CFI conflicts with the control-flow behavior of real-world code.

\subsection{Limitations and Effectiveness of CFI}\label{sec:cfi_limitations}
Extensive work analyzes CFI's residual attack surface and bypasses. 
Over-approximate target sets allow control-flow bending~\cite{carlini2015control}, while advanced ROP/JOP attacks remain feasible under fine-grained policies
\cite{goktas2014out,conti2015losing,evans2015control,li2020finding,davi2014hardware,christoulakis2016hcfi,zhang2015control}. 
The difficulty of constructing sound CFGs further limits enforcement precision~\cite{tan2017cfg}. 
These limitations underscore that practical security depends not only on policy strictness (the granularity of the CFI policy) but also on \textit{feasible deployment}. 
By automating the resolution of real-world compatibility issues, \tool{} addresses this underexplored requirement and enables robust application of compiler-based CFI in complex software.

\section{Conclusion}
This paper introduced \tool{}, the first fully automated framework for enabling compiler-based forward-edge Control-Flow Integrity (CFI) in large, legacy C and C++ software systems. 
While existing CFI mechanisms offer strong protection, their practical deployment remains hampered by visibility constraints, type inconsistencies, and semantic incompatibilities that surface only at build or test time. 
\tool{} addresses these challenges through a principled, iterative feedback loop that combines link-time analysis, ptrace-based runtime monitoring, and a structured escalation strategy for repairing CFI-induced failures.

Evaluated on four real-world GNU projects, \tool{} automatically resolved all visibility-related build errors and repaired 46 out of 48 unintended CFI policy violations in \texttt{util-linux}, the most structurally diverse subject in our study. 
Across all projects, the resulting builds retain strict protection for the vast majority of indirect control-flow sites; in \texttt{util-linux}, 86.29\% remain fully protected after escalation, with only 3.17\% ignored due to irreconcilable structural patterns. 
These results demonstrate that strict, type-based CFI is compatible with substantial portions of mature C code bases, provided that enforcement is supported by automated, localized repair.

More broadly, \tool{} lowers the practical barrier for adopting compiler CFI by eliminating the considerable manual effort previously required to diagnose and correct compatibility issues. 
As memory-unsafe legacy software continues to underpin critical infrastructure, automated hardening mechanisms such as \tool{} offer a realistic path toward increasing the deployment of strong, compiler-based mitigations in production environments.

\section{Data Availability}\label{sec:data}
All artifacts required to reproduce our results are publicly available. 
We release the full implementation of \tool{}, including the analysis engine, report generator, and all instrumentation components. 
We additionally provide the configuration files for the four evaluated projects, together with the corresponding reports generated in our study. 
All materials can be accessed at: \url{https://github.com/software-engineering-and-security/cfighter.git}.

\bibliographystyle{acm}
\bibliography{bib}

\appendix
\section{\texttt{ptrace}-Based Stack Unwinding}\label{app:ptrace-unwinding}

This appendix provides additional detail on the low-level mechanisms used by \tool{} to reconstruct execution state at the moment of a CFI violation.

\subsection{Register Acquisition}
The Linux \texttt{ptrace} API exposes the full general-purpose register set for a stopped thread via:

\[
\texttt{ptrace(GETREGS, pid, \ldots)}.
\]

This call provides:
\begin{itemize}
    \item the architectural program counter (\texttt{RIP}/\texttt{PC});
    \item the canonical frame pointer (\texttt{RBP} or \texttt{x29});
    \item the stack pointer (\texttt{RSP} or \texttt{SP});
    \item all callee-saved registers, ensuring consistency for frame-walking.
\end{itemize}

No runtime instrumentation or compiler modifications are required.

\subsection{Safe Memory Reads}
Both x86\_64 and AArch64 expose user memory through:

\[
\texttt{ptrace(PEEKTEXT, pid, address)}.
\]

This allows \tool{} to read stack words directly from the traced process.  All reads are performed with error-checked wrappers:

\[
\texttt{ret} = \text{result of } [\mathtt{FP} + 8].
\]

\subsection{Unwinding Invariants}

\tool{} relies on three invariants for correctness:

\begin{enumerate}
    \item \textbf{Frame-pointer integrity.}  
          Both x86\_64 and AArch64 ABIs maintain FP-chains by default when fame-pointer omission (FPO) is disabled during compilation.  
          \tool{} enforces FPO-off during instrumentation.

    \item \textbf{Monotonic frame chain.}  
          A valid frame satisfies:
          \[
            \mathrm{FP}_{n+1} > \mathrm{FP}_{n},
          \]
          preventing infinite loops or corrupt-chain traversal.

    \item \textbf{Two-level bound.}  
          To avoid unwinding corrupted frames or compiler-generated trampolines,           the walker limits itself to:
          \[
            \text{depth} \leq 2.
          \]
\end{enumerate}

\subsection{Comparison to Full DWARF Unwinders}

Unlike DWARF-based unwinders (e.g., \texttt{libunwind}, \texttt{gdb}), \tool{} intentionally performs:

\begin{itemize}
    \item \textbf{no CFA evaluation}, i.e., no computation of the \textit{Canonical Frame Address} — the abstract reference point used by DWARF unwinders to describe the logical base of each stack frame; 
    \item \textbf{no parsing of FDE/CIE tables}, i.e., no interpretation of \textit{Frame Description Entries} or \textit{Common Information Entries}, DWARF records that encode the function-specific and shared rules for reconstructing caller registers during stack unwinding;
    \item \textbf{no rule-based register reconstruction}, i.e., no use of DWARF unwinding opcodes (such as \texttt{DW\_CFA\_*}) to rebuild the full caller context.
\end{itemize}

This dramatically reduces complexity and ensures reliability even when:

\begin{itemize}
    \item the binary lacks unwind information,
    \item sanitizer runtime state is partially corrupted,
    \item the violation occurs in an inline or tail-call site.
\end{itemize}

The minimalist approach is therefore better suited for real-world CFI debugging, where speed and resilience outweigh full unwind fidelity.

\section{Architectural Differences Between x86\_64 and AArch64}
\label{app:isa_differences}

This subsection summarizes the architectural properties of x86\_64 and AArch64 that are relevant to \tool{}'s CFI-violation handling and call-chain reconstruction. The discussion remains concise while maintaining academic precision.

\subsubsection*{Exception-Generating Instructions and Program Counter Semantics}

\textbf{x86\_64.}
LLVM may signal CFI failures via an invalid opcode (\texttt{SIGILL}) or a software breakpoint (\texttt{int3}, yielding \texttt{SIGTRAP}). 
On Linux, \texttt{SIGTRAP} reports a program counter that has advanced past the one-byte \texttt{int3} instruction. 
Accordingly, \tool{} applies a program-counter correction of \(\text{RIP} - 1\) when diagnosing breakpoint-based violations. 
In contrast, \texttt{SIGILL} reports the address of the faulting instruction directly and requires no adjustment.

\textbf{AArch64.}
Clang emits the architecturally reserved \texttt{udf \#imm} instruction for CFI checks, producing \texttt{SIGILL}. 
Debug traps use \texttt{brk \#imm}, raising \texttt{SIGTRAP}. 
In both cases, the program counter recorded in the exception syndrome matches the trapping instruction exactly, and no architectural correction is necessary.

\subsubsection*{Indirect Control-Flow Instructions}

\textbf{x86\_64} performs indirect calls and jumps using register-indirect variants of \texttt{call}, \texttt{jmp}, and through return-address dispatch.  
\textbf{AArch64} uses \texttt{blr} (branch with link) and \texttt{br} (branch), yielding a more uniform indirect-branch model. 
As a result, the mapping between instrumentation sequences and control-transfer sites is often more regular on AArch64.

\subsubsection*{Stack-Frame and Unwind Conventions}

Under the System~V AMD64 ABI, stack-frame reconstruction relies on optional frame-pointer chains via \texttt{RBP}, where the saved return address is stored at \([\texttt{RBP} + 8]\). 
Compiler optimizations may omit this convention, reducing reliability in the absence of DWARF unwind metadata.

In contrast, the AAPCS64 ABI prescribes a fixed and canonical frame structure using \texttt{FP} (\texttt{x29}) and the link register \texttt{LR} (\texttt{x30}). 
The 16-byte frame record and mandatory linkage conventions facilitate consistent call-chain reconstruction, enhancing the stability of \tool{}'s escalation logic on AArch64 binaries.

\subsubsection*{Implications for \tool{}}

Despite these ISA differences, the classification and escalation pipeline employed by \tool{} remains architecture-agnostic. 
Only two components require ISA-specific handling:

\begin{enumerate}
    \item Fault-site recovery, due to differing trap and PC semantics.
    \item Frame-pointer traversal, reflecting the ABI-mandated frame structures of the respective architectures.
\end{enumerate}

All remaining analysis, interpretation, and reporting logic is shared between the x86\_64 and AArch64 implementations.

\section{Example of Compiler-Lowered Jump Tables}
\label{app:lowered-jt}

Modern compilers frequently optimize high-level \texttt{switch} statements by translating them into low-level indexed dispatch code. 
This transformation eliminates the structured \texttt{switch} instruction in the LLVM intermediate representation (IR). 
It replaces it with a table of code pointers or block addresses together with a computed indirect branch or call. 
In the context of forward-edge CFI, the two forms differ substantially: 
the high-level IR instruction retains explicit multi-way control-flow structure, whereas the lowered form behaves similarly to an untyped function-pointer dispatch. 
The latter is more difficult for type-based CFI mechanisms to classify precisely and therefore constitutes an essential source of compatibility challenges.

\paragraph{High-Level Form.}
The following C function illustrates a typical multi-way branch.
\begin{figure}[!ht]
\centering
\begin{lstlisting}[style=code,language=C++]
int f(int x) {
    switch (x) {
    case 0: return 10;
    case 1: return 20;
    case 2: return 30;
    default: return -1;
    }
}
\end{lstlisting}
\caption{Code example of \textit{high-level} jump table (\texttt{switch})}
\label{lst:jp_switch}
\end{figure}

Without optimizations, Clang emits a structured \texttt{switch} instruction in LLVM IR:

\begin{figure}[!ht]
\centering
\begin{lstlisting}[style=code,language=IR]
switch i32 %x, label %default [
  i32 0, label %case0
  i32 1, label %case1
  i32 2, label %case2
]
\end{lstlisting}
\caption{Code example of jump table (\texttt{switch}) translated to LLVM IR}
\label{lst:ir_switch}
\end{figure}

This representation exposes all outgoing edges directly and preserves the multi-way control-flow structure that CFI can interpret with high precision. 
Such instances correspond to ``JT (switch)'' in our static analysis. 

\paragraph{Lowered Form.}
At most optimization levels, LLVM replaces a structured branch with a jump table. 
The optimized IR materializes a static table of code pointers and performs a computed indirect call: 
\begin{figure}[!ht]
\centering
\begin{lstlisting}[style=code,language=IR]
%table = constant [3 x i32 (...)*] [
    i32 (...)* @f.case0,
    i32 (...)* @f.case1,
    i32 (...)* @f.case2
]

%idx  = load i32, i32* %x.addr
%ptr  = getelementptr [3 x i32 (...)*],
                [3 x i32 (...)*]* @table, i32 0, i32 %idx
%func = load i32 (...)*, i32 (...)** %ptr
%res  = call i32 %func()
\end{lstlisting}
\caption{Code example of \textit{lowered} jump table translated to LLVM IR}
\label{lst:jp_lowered}
\end{figure}

In this lowered form, the original \texttt{switch} structure is no longer explicit. 
The dispatch is implemented as a computed pointer lookup followed by an indirect call, which appears to the CFI mechanism as a generic function pointer call rather than as a multi-way branch with known alternatives. 
These instances correspond to ``JT (lowered)'' in our analysis. 

This transformation is benign for performance but has important implications for CFI: 
the lack of structural metadata and the introduction of computed indirection reduce the precision of static type-based checks, making lowered jump tables an important factor in the design and evaluation of \tool{}.

\end{document}